\documentclass[pra,10pt,superscriptaddress,footnoteinbib]{revtex4}
\usepackage{amsmath}
\usepackage{latexsym}
\usepackage{amssymb}
\usepackage{graphics,epstopdf}
\usepackage[colorlinks=true, citecolor=blue, urlcolor=blue]{hyperref}\usepackage{epsf,graphics,graphicx}

\newcommand{\n}{\noindent}

\newcommand{\ed}{\end{document}}
\newcommand{\beq}{\begin{equation}}
	\newcommand{\eeq}{\end{equation}}

\begin{document}\title{Effect of Perturbative Hexagonal Warping on Quantum Capacitance in Ultra-Thin Topological Insulators}
	\author{Anirudha Menon}
	\affiliation{Department of Physics, University of California, Davis, California 95616, USA}\email{amenon@ucdavis.edu}
	\author{Debashree Chowdhury}
	\affiliation{Department of Physics, Indian Institute of Technology, Kanpur 208016, India}\email{debashreephys@gmail.com}
	\author{Banasri Basu}
	\affiliation{Physics and Applied Mathematics Unit, Indian Statistical Institute, Kolkata 700108, India}
	\email{sribbasu@gmail.com}

\graphicspath{{C:/images/}}
%%\vspace*{1cm}

\begin{abstract}
\n
Ultra-thin 3D topological insulators provide a stage to study the surface physics of such materials by minimizing the bulk contribution. Further, the experimentally verified snowflake like structure of the Fermi surface leads to a hexagonal warping term, and this discourse examines it as a perturbation in the presence of a magnetic field. We find that there are corrections to both energy dispersion and eigenstates which in turn alter the density of states in the presence of a magnetic field. Both the quantum capacitance and the Hall coefficient are evaluated analytically and it is shown here that we recover their established forms along with small corrections which preserve the object of treating hexagonal warping perturbatively. In our approach, the established Hall conductivity expression develops several minute correction terms and thus its behavior remains largely unaffected due to warping. The zero-temperature quantum capacitance exhibits SdH oscillations with reduced frequencies, with a lowered average capacitance with increased warping of the Fermi surface, while maintaining the usual amplitudes. 
\end{abstract}

\maketitle
~~~~~PACS numbers: 73.20.-r, ~75.70.Tj,~73.43.Cd

\maketitle
\section{Introduction}
Recent studies on a new state of matter called a topological insulator (TI) have shown potential advancement
for spintronic applications~\cite{SpinA1, SpinD1, SpinD2} due to an inherent spin-momentum locking described by the Rashba~\cite{Rash}
Hamiltonian, characterized by a Dirac piece. Materials like $Bi_{2}Te_{3}$, $Sb_{2}Te_{3}$, and $Bi_{2}Se_{3}$ have been known to exhibit the properties of a TI~\cite{Mater}, which involves conduction of electrons on the surface of an insulator. Scanning tunneling microsopy~\cite{SCM} and photoemission spectroscopy~\cite{PES1, PES2, PES3} experiments have shown the existence of the single Dirac cone in TI's, verifying the proposed linear energy spectrum.  However, as most theoretical insulators are in practice conductors due to doping and 
crystal lattice defects, transport phenomena are difficult to map accurately, and to this end, research is 
now directed towards thin film TI's which facilitate the exclusion of contributions from the bulk~\cite{UTI1,UTI2,UTI3}. Essentially, for some 3D TIs (e.g. $Bi_{2}Te_{3}$), when one increases chemical potential beyond the Dirac point, a continuous change in the Fermi surface can be obtained  due to helical Dirac Fermions~\cite{DFerm1,Dferm2} existing on the surface. The shape of the corresponding Fermi surface changes from a circle to a snowflake like structure~\cite{Snow}. This phenomenon is well described by a hexagonal warping term proposed by Fu~\cite{Fu} in addition to the original Rashba Hamiltonian. It is worth noting that this new Hamiltonian is 3-fold invariant while the corresponding energy dispersion is 6-fold invariant due to 
time reversal symmetry~\cite{Fu}. 

More significantly, in the ultra-thin limit the top and bottom states of a TI begin to hybridize~\cite{Hybd1,Hybd2}
leading to a cross over from 3D to 2D surface states. This is easily accounted for by adding a small
hybridization term to the TI Hamiltonian and it can be shown that Schr\"odinger 
piece, which is the generator of particle-hole asymmetry~\cite{Schro}, is negligible in this context. 
Properties of ultra-thin TI's such as magneto-optical response~\cite{MO1, MO2} and quantum spin Hall~\cite{QSH2} and anomalous 
Hall~\cite{QAH} effects, and possible exitonic super-fluidity~\cite{SuperF} have been studied theoretically, along with spin texture and circular dichroism in the absence of a magnetic field~\cite{Zhou}. A literature search reveals that most theoretical studies in this this field have been conducted assuming a uniform Fermi surface and so it is useful to study the effects of the inhomogeneity induced by the hexagonal warping term on different physical quantities. We consider the impact of the warping term on the Hall conductivity and the zero temperature quantum capacitance in ultrathin TI's, and illucidate on the characteristics of these quantities which are different from the non-warping case and maybe experimentally verifiable.

 However, the exact solution to the Hamiltonian in its gauge invariant form in the presence of a magnetic 
field as well as warping is likely mathematically intractable as are most an-harmonic oscillators. To our advantage, we treat the hexagonal warping parameter $\lambda$ to be small~\cite{Zhou} with the Dirac piece being the dominant contribution up to the Fermi surface. As a consequence, in this paper we 
will be looking at the effects of hexagonal 
warping on Hall conductivity and quantum capacitance in the presence of a magnetic field using time independent perturbation theory. The corrections to the energy dispersion and states of the Rashba Hamiltonian with the 
Zeeman and hybridization pieces will be determined treating the hexagonal warping term perturbatively. 

The following is the structure of the paper. In section II we reiterate the solution to the Rashba Hamiltonian along with the Zeeman and hybrization terms in the presence of a magnetic field~\cite{Burk}, ignoring the Schr\"odinger term. We then consider the hexagonal warping term perturbatively and we derive the first and second order corrections to the energy dispersion. We present an analytical calculation for the asymptotic density of states (DOS) in the presence of a magnetic field using the technique of self energy. In section IV we determine the correction to the eigenstates of the Hamiltonian stated above using first order perturbation theory and use this result to evaluate the Hall coefficient. Quantum capacitance is determined analytically in Section V using the DOS obtained in Section III. Conclusions are presented in Section VI.

\section{The ultrathin TI  Hamiltonian and the Warping Spectrum}
We begin by considering a Hamiltonian of a TI, similar to the one proposed by Fu, but with an additional piece to account for the hybridization of the top and bottom states in the ultra-thin regime~\cite{Zhou}.
\beq H = \hbar v_{F}(\sigma_{x}\pi_{y} - \tau_{z}\sigma_{y}\pi_{x}) +(\Delta_z \sigma_z + \Delta_t \tau_x) + \frac{\lambda}{2}\sigma_{z}(\pi_{+}^{3} + \pi_{-}^{3}),\label{1}\eeq

The first term in the r.h.s of (\ref{1}) is due to spin-momentum locking effect, with $v_{F}$ as the Fermi velocity. $\sigma_{i}$ and $\tau_{i}$  with $i = x, y, z$ are thel spin and the surface pseudospin degrees of freedom. Here $\pi$ denotes the minimal momentum and $\tau_{z} =\pm 1$ denotes symmetric and anti-symmetric
surface states respectively. The second term has two components, the first of which is for Zeeman interaction $\Delta_{z} = \frac{g\mu_{B}\vec{B}}{2}$, arising because of the external magnetic field $\vec{B} = B\hat{z}$ with $g$ and $\mu_{B}$ being  the electron g factor and Bohr magneton respectively. The second component  $\Delta_{t}$ is the hybridization energy and it accounts for the tunneling between the two surface states. The last term in the right hand sight of (1) is the hexagonal warping term, with $\lambda$ as the warping parameter. 
 We denote $H=H_{0}+ H_{p}$ where 
  $H_{0} = \hbar v_{F}(\sigma_{x}\pi_{y} - \tau_{z}\sigma_{y}\pi_{x}) +\Delta_z \sigma_z + \Delta_t \tau_x $ and $H_{p} = \frac{\lambda}{2}\sigma_{z}(\pi_{+}^{3} + \pi_{-}^{3})$ for convenience. We ignore the warping term for the moment and proceed to state the solution to $H_{0}$. In order to diagonalize the Hamiltonian $H_{0}$, Landau operators are introduced as 
$a = \frac{l}{\sqrt{2}}(\pi_{x} - i\pi_{y}),~~a^{\dag} = \frac{l}{\sqrt{2}}(\pi_{x} + i\pi_{y}),$ where $l = \sqrt{\frac{c}{eB}}$ is the magnetic length and $\vec{\pi} = -i\vec{\nabla} + \frac{e}{c}\vec{A},$ is the gauge invariant momentum. In terms of the ladder operators $H_{0}$ is reexpressed as 
\beq\label{2}
H_{0} = \frac{i \omega_c}{\sqrt{2}} \tau_z (\sigma_+ a - \sigma_- a^\dag) + \Delta_z \sigma_z + \Delta_t \tau_x,\eeq
where $\omega_c = v_F / \ell,$ is the characteristic frequency analogous to the cyclotron frequency of a usual 2DEG. 

The single-particle eigenstates have the following general form~\cite{Burk}:
\begin{eqnarray}
\label{3}
| n \alpha  s \rangle = u^{\alpha  s}_{n T \uparrow} | n-1, T, \uparrow \rangle + u^{\alpha s}_{n T \downarrow} | n, T, \downarrow \rangle  + u^{\alpha  s}_{n B \uparrow} | n-1, B, \uparrow \rangle + u^{\alpha s}_{n B \downarrow} | n, B, \downarrow \rangle.
\end{eqnarray}
$| n, T(B), \uparrow(\downarrow) \rangle$ represents the $n$-th Landau Level (LL) eigenstate~\cite{Burk} pertaining to the Hamiltonian in Eq.(\ref{2}), on the top (bottom) surface with spin up (down). Here 
$\alpha=0, 1$, $s = \pm$, $n = 0,\ldots,\infty$, and 
$u^{\alpha  s}_n$ are the corresponding complex coefficients, which can be obtained as
\beq
\label{4}
u^{\alpha s}_n = \left[ i s (-1)^{\alpha}  f_{n \alpha s +}, -s f_{n \alpha s -}, i (-1)^{\alpha} f_{n \alpha s +}, f_{n \alpha s -} \right],  
\eeq
where
\beq
\label{5}
f_{n \alpha s \pm} = \frac{1}{2} \sqrt{1 \pm \frac{\Delta_z + s \Delta_t}{\epsilon_{n \alpha s}}}.
\eeq
The corresponding state energy is:
\beq
\label{6}
E_{n \alpha s} = (-1)^{\alpha}\sqrt{2 {\omega^2_{c}} n +({\Delta}_z + s{\Delta}_t)^2}.
\eeq

Since the complete Hamiltonian also contains an anharmonic term $H_p$, we proceed to the use of time independent perturbation theory to determine the corresponding energy corrections. Only the dynamical component of the ladder operators will be utilized for this purpose as it can be readily shown that the contribution from the field is negligible. The first order correction to energy collapses trivially as 
\beq E_{n \alpha s}^{1} = \langle n\alpha s|\frac{\lambda}{2}\sigma_{z}(\pi_{+}^{3} + \pi_{-}^{3})|n\alpha s\rangle = 0.\eeq  
As is standard, we therefore proceed with second order energy correction calculations, and the general form for this is as follows. 
\beq  E_{n \alpha s}^{2} = \sum_{n^{'} \alpha^{'} s^{'} \neq n \alpha s}\frac{|<n\alpha s|H_{p}|n^{'}\alpha^{'}s^{'}>|^{2}}{E_{n} - E_{n^{'}}}.\eeq
Note that the summations in the above expression are only on the primed variables. A little algebraic manipulation on the determinant of the matrix element $<n\alpha s|H_{p}|n^{'}\alpha^{'}s^{'}>,$ yields

\begin{eqnarray} 
|<n\alpha s|H_{p}|n^{'}\alpha^{'}s^{'}>|&=& A_{n^{'}+2}u^{\alpha  s *}_{n T \uparrow}u^{\alpha^{'}  s^{'}}_{n^{'} T \uparrow}<n-1, T,\uparrow|n^{'}-2, T,\uparrow \rangle - A_{n^{'}+3}u^{\alpha  s *}_{n T \downarrow}u^{\alpha^{'}  s^{'}}_{n^{'} T \downarrow}<n, T,\downarrow|n^{'}+3, T,\downarrow \rangle\\ &+&  A_{n^{'}+2}u^{\alpha  s *}_{n B \uparrow}u^{\alpha^{'}  s^{'}}_{n^{'} B \uparrow}<n-1, B,\uparrow|n^{'}+2, B,\uparrow \rangle - A_{n^{'}+3}u^{\alpha  s *}_{n B \downarrow}u^{\alpha^{'}  s^{'}}_{n^{'} B \downarrow}<n, B,\downarrow|n^{'}+3, B,\downarrow \rangle \nonumber\\ &+& A_{n^{'}-3}u^{\alpha  s *}_{n T \uparrow}u^{\alpha^{'}  s^{'}}_{n^{'} T \uparrow}<n-1, T,\uparrow|n^{'}-4, T,\uparrow \rangle - A_{n^{'}-2}u^{\alpha  s *}_{n T \downarrow}u^{\alpha^{'}  s^{'}}_{n^{'} T \downarrow}<n, T,\downarrow|n^{'}-3, T,\downarrow \rangle\nonumber\\ &+&
A_{n^{'}-3}u^{\alpha  s *}_{n B \uparrow}u^{\alpha^{'}  s^{'}}_{n^{'} B \uparrow}<n-1, B,\uparrow|n^{'}-4, B,\uparrow \rangle - A_{n^{'}-2}u^{\alpha  s *}_{n B \downarrow}u^{\alpha^{'}  s^{'}}_{n^{'} B \downarrow}<n, B,\downarrow|n^{'}- 3, B,\downarrow \rangle \nonumber,
\end{eqnarray}
where $A_{n^{'}+p} = \sqrt{(n^{'}+p)(n^{'}+p-1)(n^{'}+p - 2)}$ and $A_{n^{'} -p} = \sqrt{(n^{'}-p)(n^{'}-p-1)(n^{'}-p - 2)}.$ \\

For a non-zero result, the conditions $\langle n|n^{'}\rangle = \delta_{n n^{'}} \neq 0$ and $s=s^{'}$ are mandatory. On explicitly writing out each term and substituting in the second order energy correction formula,  it is easy to arrive at the following expression. 

\begin{equation} 
\begin{split}
E_{n \alpha s}^{2} =\frac{\lambda^{2}}{16}\sum_{\alpha^{'}}\left\{\frac{(n-2)(n-1)}{E_{n\alpha s} - E_{n-3\alpha^{'}s}}\left[\sqrt{n-3}(-1)^{\alpha+\alpha^{'}}(1+\frac{\Delta_{z}+s\Delta_{t}}{E_{n\alpha s}})^{\frac{1}{2}}(1+\frac{\Delta_{z}+s\Delta_{t}}{E_{n-3\alpha^{'} s}})^{\frac{1}{2}}\right. \right.\\ \left.\left. - \sqrt{n}(1-\frac{\Delta_{z}+s\Delta_{t}}{E_{n\alpha s}})^{\frac{1}{2}}(1-\frac{\Delta_{z}+s\Delta_{t}}{E_{n-3\alpha^{'} s}})^{\frac{1}{2}}\right]+ \frac{(n+2)(n+1)}{E_{n\alpha s} - E_{n+3\alpha^{'}s}}\left[\sqrt{n}(-1)^{\alpha+\alpha^{'}}(1+\frac{\Delta_{z}+s\Delta_{t}}{E_{n\alpha s}})^{\frac{1}{2}}(1+\frac{\Delta_{z}+s\Delta_{t}}{E_{n+3\alpha^{'} s}})^{\frac{1}{2}}\right. \right.\\ \left.\left. - \sqrt{n+3}(1-\frac{\Delta_{z}+s\Delta_{t}}{E_{n\alpha s}})^{\frac{1}{2}}(1-\frac{\Delta_{z}+s\Delta_{t}}{E_{n+3\alpha^{'} s}})^{\frac{1}{2}}\right]\right\}
\end{split}
\end{equation}
The complete energy spectrum $E^T_{n \alpha s}$  is therefore
\beq E_{n \alpha s}^{T}= E_{n \alpha s} + E_{n \alpha s}^2 \eeq

\begin{figure}[h]
\centering
\begin{minipage}[b]{0.45\linewidth}
\fbox{\includegraphics[scale=0.6]{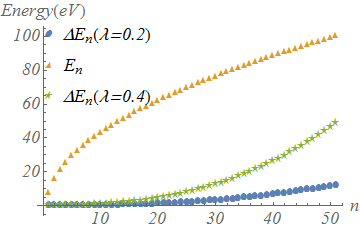}}
\caption{Perturbative energy correction in comparison to LL spectrum with $\lambda = 0.2$ and $0.4$ eV $nm^3$, $\alpha=0$ and $s=1$.}
\label{fig:minipage1}
\end{minipage}
\quad
\begin{minipage}[b]{0.45\linewidth}
\fbox{\includegraphics[scale=0.6]{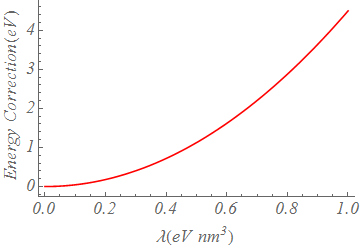}}
\caption{A plot of second order energy corrections for different values of warping parameter in the range [0,1].}
\label{fig:minipage1}
\end{minipage}
\end{figure}

Fig.1 shows the plot between the energy (the unperturbed energy and the second order correction to the energy) on the $y$ axis with the Landau Level index $n$ on the $x$ axis. One can see that the energy corrections for different values of $\lambda$ are small, which justifies $\lambda$ being treated perturbatively within the limits of the Fermi surface. Here, we have assumed the following~\cite{Burk,Tahir}: B= 10 T, $\Delta_z = 5$ meV, $\Delta_t$ = 3 meV, and these parameters remain unaltered throughout unless stated otherwise. 
Although for a given $\lambda$ there may exist an $n$ for which $E_n < E_n^2$, this breakdown cannot be realised in any physical material as the required occupation of LL's far exceeds that of known materials~\cite{Fu}. Fig.2 depicts that the second order correction to energy due to perturbation increases quadratically with $\lambda$, i.e. the warping parameter. We will utilize the energy correction derived above in the subsequent sections.

\section{Density of states calculation}
This section contains the formulation of the density of states considering the warped Fermi surface in a perpendicular magnetic field. DOS calculations are a fundamental aspect of material science which facilitate the calculation of the Hall conductivity and the longitudinal conductivity amongst others. There's a strong correlation between the DOS and the zero temperature quantum capacitance which we shall see in Section V. Here we proceed using the technique of self energy.

Considering the self energy of the system as $\Sigma^{-}(E)$, one can write the following recursive relation~\cite{SE1,SE2} \beq \Sigma^{-}(E) = \Gamma_{0}^{2}\sum_{n}\frac{1}{E -E_{n}^T-\Sigma^{-}(E)}, \label{18}\eeq $\Gamma_0$ is the impurity induced LL broadening in eqn(\ref{18}). The DOS is then expressed in terms of self energy as \beq D(E) = Im[\frac{\Sigma^{-}(E)}{\pi^{2}l^{2}}\Gamma_{0}^{2}]. \eeq Using the perturbative energy correction derived in the previous section, the modified energy of the system is $E_{n}^{T} = E_{n} + E_{n}^{2}$, where we've dropped the labels $\alpha$ and $s$. Drawing the  conclusion that $E_{n}^{2} << E_{n}$, and introducing it in eqn (\ref{18}), we arrive at  \beq \Sigma^{-}(E) = \Gamma_{0}^{2}\sum_{n}\frac{1}{E -E_{n}-\Sigma^{-}(E)} + \Gamma_{0}^{2}\sum_{n}\frac{E_{n}^{2}}{[E -E_{n}-\Sigma^{-}(E)]^{2}},\eeq on Taylor expansion. The form of this expression is of great advantage as we are familar with the closed form for the first term~\cite{Firoz1} as shown below. 
\beq \Gamma_{0}^{2}\sum_{n}\frac{1}{E -E_{n}-\Sigma^{-}(E)} = \frac{2\pi \Gamma_{0}^{2} E}{(\hbar w_{c}^{2})^{2}} cot(\pi n_{0}),\eeq where $n_{0} = \frac{1}{(2\hbar w_{c}^{2})^{2}}[\{E-\Sigma^{-}(E)\}^{2} - (\Delta_{z}+s\Delta_{t})^{2}]$ is the pole. \\
\\
The above sum incorporates a factor of two due to the symmetric filling of states for either value of $\alpha$, and we will extend this to the algebra pertaining to the second term. This term shares a repeated pole with the first term and we utilise the residue theorem and a standard summation formula in its evaluation. Then,

\beq \Sigma^{-}(E)_{second~term} = -\frac{\pi \Gamma_{0}^{2}}{4(\hbar w_{c}^{2})^{4}}\left[4E E^{2}_{n} cot (\pi n)  + 4E^{2} E_{n}^{2} \left[-\pi cosec^{2}(\pi n)  + cot(\pi n) \frac{d}{dn} E^{2}_{n} \right]\right]_{n = n_{0}}.\eeq
Noting that $E\ll 1,$ neglecting higher order terms in E and introducing the factor due to $\alpha$, lands us at

\beq \Sigma^{-}(E) = \frac{2\pi \Gamma_{0}^{2}E}{(\hbar w_{c})^{2}}\left(1-\frac{ E_{n_0}^2}{(\hbar w_{c})^{2}}\right)cot \pi n_{0}.\eeq
Ignoring the self energy contribution to the pole, and using the method in ref.[3] we can extract the total DOS by summing over each $s$ branch. 
\beq D(E) = \sum_{s=\pm 1}\frac{D_{0}(E)}{2}\left [1+2\sum_{u=1}^{\infty} exp\{-u(\frac{2\pi\Gamma_{0}E}{\hbar^{2}w_{c^{2}}})^2\}cos\{\frac{u\pi({E}^2 - (\Delta_z + s \Delta_t)}{(\hbar w_{c})^{2}}\}\right]\left(1-\frac{ E^2_{n_{0}}}{(\hbar w_{c})^{2}}\right),\label{22}\eeq
where $n_{0} = \frac{1}{(\hbar w_{c})^{2}}\left[E^{2} - (\Delta_{z} +s \Delta_{t})^{2}\right].$ \\
\\
 We've recovered the asymptotic form ($n>1$) of the DOS~\cite{Firoz1, Tahir} with a minute additional term which manifests due to warping. It turns out that the above expression can be further simplified by reexpressing the infinite sum as follows:
\\
\beq S =  Re \left [\quad \sum_{u=0}^{\infty} exp{\{-\gamma^2 u +i \beta u \}}  \right]  \eeq
with
\beq \gamma=(\frac{2\pi \Gamma_0 E}{\hbar^2  w_c^2}) \quad , \quad \beta =(\pi \frac{[{E}^2 - (\Delta_z + s \Delta_t)^2]}{\hbar^2 w_c^2}) .\eeq
Summing the infinite geometric series we obtain,
\beq S= Re\left[\frac{exp\{\gamma^2 \}}{exp\{\gamma^2 \} - exp\{i\beta \}}\right] = \frac{1 - exp\{-\gamma^2\}Cos\beta}{1 - 2exp\{-\gamma^2\}Cos\beta + exp\{-2\gamma^2\}} .\eeq
The final expression for the DOS in the presence of a magnetic field then reads
\beq D(E) = D(E,1) + D(E,-1) ,\eeq
with $ D(E,s)$  defined as
\beq D(E,s) = \frac{{D_0}(E)}{2} \left[ \frac{1 - exp\{-(\frac{2\pi \Gamma_0 E}{\hbar^2  w_c^2})^2\}Cos(\pi \frac{[{E}^2 - (\Delta_z + s \Delta_t)^2]}{\hbar^2 w_c^2})}{1 - 2exp\{-(\frac{2\pi \Gamma_0 E}{\hbar^2  w_c^2})^2\}Cos(\pi \frac{[{E}^2 - (\Delta_z + s \Delta_t)^2]}{\hbar^2 w_c^2}) + exp\{-2(\frac{2\pi \Gamma_0 E}{\hbar^2  w_c^2})^2\}}\right] \left(1-\frac{ E^2_{n_{0}}}{(\hbar w_{c})^{2}}\right) \eeq

\begin{figure}[h]
\centering
\begin{minipage}[b]{0.45\linewidth}
\fbox{\includegraphics[scale=0.6]{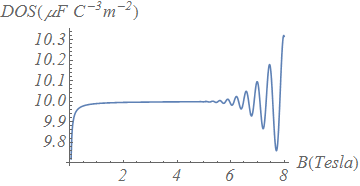}}
\caption{Density of states plotted against magnetic field varying in the range 0.1T to 8T}
\label{fig:minipage1}
\end{minipage}
\end{figure}
Fig. 3 shows how the DOS varies from $B=0.1T$ to $B=8$T and one can see that SdH oscillations set in around 5.5T which may be a consequence of the asymptotic form of the DOS. Also, the function is singular at $B=0$ and so it cannot be used to recover the DOS in the absence of a magnetic field. This is typical of quantum systems where the removal of a quantizing field by setting it equal to zero does not yield desired results. Since, in this paper we are dedicated to results in the presence of the magnetic field, we will ignore the behavior of the DOS in the absence of a magnetic field. In the next two sections we will discuss about the role of the above calculated DOS on different physical quantities.

\section{Hall conductivity}
The Hall coefficient of any conducting material is of interest to condensed matter physics~\cite{HallC, HallC2}, and to this end, we present the following calculation which reveals an analytic form for the Hall coefficient of an ultra-thin TI.
To determine this,we apply the Kubo formula for which one requires the perturbatively corrected eigenstates of the Hamiltonian presented in section II. The form of the first order eigenstate correction is 
\beq |\psi^n_{corrected} \rangle = |\psi^{n}\rangle + \sum_{n^{'} \neq n}\frac{|<\psi^{n^{'}}|H_{p}|\psi^{n}>|}{E_{n} - E_{n^{'}}}|\psi^{n^{'}}>.\eeq
Since the calculation here involving the evaluation of the matrix element $<\psi^{n^{'}}|H_{p}|\psi^{n}>$ bears a stark similarity to the energy correction, we'll state the result directly, but we note that the equation above needs to be modified with sums over $\alpha^{'}$ and $s^{'}$ as a starting point and as usual the requirement $s=s^{'}$ holds.

\begin{eqnarray}
|n~\alpha~s\rangle &=& |n\alpha s\rangle_{0} + \frac{\lambda}{2}\sum_{\alpha^{'}}\frac{\left[A_{n-3}u^{\alpha  s *}_{n T \uparrow}u^{\alpha^{'}  s}_{n-3 T \uparrow} - A_{n-2}u^{\alpha  s *}_{n T \downarrow}u^{\alpha^{'}  s}_{n-3 T \downarrow} + A_{n-3}u^{\alpha  s *}_{n B \uparrow}u^{\alpha^{'}  s}_{n-3 B\uparrow} - A_{n-2}u^{\alpha  s *}_{n B \downarrow}u^{\alpha^{'}  s}_{n-3 B \downarrow}\right]}{E_{n} - E_{n-3}}|n-3~\alpha ~s\rangle_{0}\nonumber\\
&+& \frac{\lambda}{2}\sum_{\alpha^{'}}\frac{\left[A_{n+3}u^{\alpha  s *}_{n T \uparrow}u^{\alpha^{'}  s}_{n+3 T \uparrow} - A_{n+2}u^{\alpha  s *}_{n T \downarrow}u^{\alpha^{'}  s}_{n-3 T \downarrow} + A_{n+3}u^{\alpha  s *}_{n B \uparrow}u^{\alpha^{'}  s}_{n+3 B\uparrow} - A_{n+2}u^{\alpha  s *}_{n B \downarrow}u^{\alpha^{'}  s}_{n+3 B \downarrow}\right]}{E_{n} - E_{n+3}}|n+3~\alpha ~s\rangle_{0}\nonumber\\
&=&  |n~\alpha~s\rangle_{0} + C_{n-3~\alpha~s}|n-3~\alpha ~s\rangle_{0} + C_{n+3~\alpha~s}|n+3~\alpha ~s\rangle_{0}, \label{sss}
\end{eqnarray}
where the $C_{n}$'s are defined as
\begin{eqnarray}
C_{n-3~\alpha~s} &=& \frac{\lambda}{2} \sum_{\alpha^{'}}\frac{\left[A_{n-3}u^{\alpha  s *}_{n T \uparrow}u^{\alpha^{'}  s}_{n-3 T \uparrow} - A_{n-2}u^{\alpha  s *}_{n T \downarrow}u^{\alpha^{'}  s}_{n-3 T \downarrow} + A_{n-3}u^{\alpha  s *}_{n B \uparrow}u^{\alpha^{'}  s}_{n-3 B\uparrow} - A_{n-2}u^{\alpha  s *}_{n B \downarrow}u^{\alpha^{'}  s}_{n-3 B \downarrow}\right]}{E_{n} - E_{n-3}}\nonumber\\
 C_{n+3~\alpha~ s} &=& \frac{\lambda}{2} \sum_{\alpha^{'}}\frac{\left[A_{n+3}u^{\alpha  s *}_{n T \uparrow}u^{\alpha^{'}  s}_{n+3 T \uparrow} - A_{n+2}u^{\alpha  s *}_{n T \downarrow}u^{\alpha^{'}  s}_{n-3 T \downarrow} + A_{n+3}u^{\alpha  s *}_{n B \uparrow}u^{\alpha^{'}  s}_{n+3 B\uparrow} - A_{n+2}u^{\alpha  s *}_{n B \downarrow}u^{\alpha^{'}  s}_{n+3 B \downarrow}\right]}{E_{n} - E_{n+3}}.
\end{eqnarray}
To clarify, the subscript zero denotes the un-corrected eigenstates. The coupling of states due to the introduction of the magnetic field removes the complete spin polarization of the $n=0$ LL~\cite{Burk}, and the Dirac Fermions don't possess a zero mode anomaly. The Kubo-formula is stated below and subsequently we insert the expression from eqn (\ref{sss}) to determine the Hall coefficient. Note that the primed variables in the summation below are dummy indices and bear no relevance to prior calculations.
\begin{eqnarray} 
\sigma_{xy}&=&\frac{\omega_B^2 e^2}{2 \pi} \sum_{n \alpha s \ne n' \alpha' s'} \textrm{Im} \left[ \langle n  \alpha s | \tau^z \sigma^y |
n'  \alpha'  s' \rangle \times \langle n'  \alpha'  s' | \tau^z \sigma^x | n  \alpha  s \rangle \right]
\frac{n_F(E_{n \alpha s}) - n_F(E_{n' \alpha' s'})}{(E_{n \alpha s} - E_{n' \alpha' s'})^2}\nonumber\\
\end{eqnarray}
Keeping only leading order correction terms in $|C_m|^2$ since they're small, and accounting for like terms,
\begin{eqnarray}
\sigma_{xy}&=& \frac{\omega_B^2 e^2}{2 \pi} \sum_{n \alpha s \ne n' \alpha' s'}   \textrm{Im} \left [\langle n  \alpha s |_{0} \tau^z \sigma^y |
n'  \alpha'  s' \rangle_{0} \times \langle n'  \alpha'  s' |_{0} \tau^z \sigma^x | n  \alpha  s \rangle_{0} \right]
\frac{n_F(E_{n \alpha s}) - n_F(E_{n' \alpha' s'})}{(E_{n \alpha s} - E_{n' \alpha' s'})^2}\nonumber\\
&+& \frac{\omega_B^2 e^2}{\pi} \sum_{n \alpha s \ne n' \alpha' s'} |C_{n-3~\alpha~s}|^2  \textrm{Im}~ \left [ \langle n-3  \alpha s |_{0} \tau^z \sigma^y |
n'  \alpha'  s' \rangle_{0} \times \langle n'  \alpha'  s' |_{0} \tau^z \sigma^x | n-3  \alpha  s \rangle_{0} \right]
\frac{n_F(E_{n \alpha s}) - n_F(E_{n' \alpha' s'})}{(E_{n \alpha s} - E_{n' \alpha' s'})^2}\nonumber\\
&+&  \frac{\omega_B^2 e^2}{\pi} \sum_{n \alpha s \ne n' \alpha' s'} |C_{n+3~\alpha~s}|^2 \textrm{Im}~ \left[ \langle n+3  \alpha s |_{0} \tau^z \sigma^y |
n'  \alpha'  s' \rangle_{0} \times \langle n'  \alpha'  s' |_{0} \tau^z \sigma^x | n+3  \alpha  s \rangle_{0} \right]
\frac{n_F(E_{n \alpha s}) - n_F(E_{n' \alpha' s'})}{(E_{n \alpha s} - E_{n' \alpha' s'})^2}. \nonumber\\
\label{sig}
\end{eqnarray}
On simplification, we get

\begin{eqnarray}
\sigma_{xy}&=& \frac{e^2}{4 \pi}  \sum_{n = 0}^{\infty} \sum_{\alpha =0,1} \sum_{s = \pm} { (2 n +1) 
[n_F(\epsilon_{n \alpha s}) - n_F(\epsilon_{n+1 \alpha s})] + (\Delta_z + s \Delta_t) [ \frac{n_F(\epsilon_{n+1 \alpha s})}{\epsilon_{n+1 \alpha s}} - 
 \frac{n_F(\epsilon_{n \alpha s})}{\epsilon_{n \alpha s}} ]} \nonumber \\
&+& \frac{\omega_B^2 e^2}{\pi} \sum_{n} \sum_{\alpha s \ne \alpha' s'} |C_{n-3~\alpha~s}|^2  \left [(s^2 {s^{'2}}+1)f^2_{n \alpha s +}f^2_{n-2~\alpha ^{'} s^{'} -} - ss^{'} (f_{n-2~\alpha^{'} s^{'} +}^2 f_{n \alpha s -}^2 - f_{n-2~\alpha^{'} s^{'} -}^2 f_{n \alpha s +}^2)  \right] \nonumber \\
& \times & \frac{n_F(E_{n \alpha s}) - n_F(E_{n-2~\alpha' s'})}{(E_{n \alpha s} - E_{n-2~\alpha' s'})^2}\nonumber\\
&+& \frac{\omega_B^2 e^2}{\pi} \sum_{n} \sum_{\alpha s \ne \alpha' s'} |C_{n-3~\alpha~s}|^2  \left [(s^2 {s^{'2}}+1)f^2_{n \alpha s -}f^2_{n-4~\alpha ^{'} s^{'} +} - ss^{'} (f_{n-4~\alpha^{'} s^{'} -}^2 f_{n \alpha s +}^2 - f_{n-4~\alpha^{'} s^{'} +}^2 f_{n \alpha s -}^2)  \right] \nonumber \\
& \times & \frac{n_F(E_{n \alpha s}) - n_F(E_{n-4~\alpha' s'})}{(E_{n \alpha s} - E_{n-4~\alpha' s'})^2}\nonumber\\
&+& \frac{\omega_B^2 e^2}{\pi} \sum_{n} \sum_{\alpha s \ne \alpha' s'} |C_{n+3~\alpha~s}|^2  \left [(s^2 {s^{'2}}+1)f^2_{n \alpha s +}f^2_{n+4~\alpha ^{'} s^{'} -} - ss^{'} (f_{n+4~\alpha^{'} s^{'} +}^2 f_{n \alpha s -}^2 - f_{n+4~\alpha^{'} s^{'} -}^2 f_{n \alpha s +}^2)  \right] \nonumber \\
& \times & \frac{n_F(E_{n \alpha s}) - n_F(E_{n+4~\alpha' s'})}{(E_{n \alpha s} - E_{n+4~\alpha' s'})^2}\nonumber\\
&+& \frac{\omega_B^2 e^2}{\pi} \sum_{n} \sum_{\alpha s \ne \alpha' s'} |C_{n-3~\alpha~s}|^2  \left [(s^2 {s^{'2}}+1)f^2_{n \alpha s -}f^2_{n+2~\alpha ^{'} s^{'} +} - ss^{'} (f_{n+2~\alpha^{'} s^{'} -}^2 f_{n \alpha s +}^2 - f_{n+2~\alpha^{'} s^{'} +}^2 f_{n \alpha s -}^2)  \right] \nonumber \\
& \times & \frac{n_F(E_{n \alpha s}) - n_F(E_{n+2~\alpha' s'})}{(E_{n \alpha s} - E_{n+2~\alpha' s'})^2}\nonumber\\
\label{sig}
\end{eqnarray}

The first term in the eqn. (\ref{sig}), denotes the contribution to the Dirac Fermion Hall conductivity for the  unperturbed Hamiltonian and has been derived apriori~\cite{Burk}. The remaining four terms are attributed to the warping correction, but are completely dominated by the first term even in the absence of Zeeman splitting and the hybridization term. The usual quantum phase transition is observed with increasing magnetic field at $E_F=0$ when $\sigma_{xy}$ jumps from zero to $\pm e^2/2\pi$.

\section{Quantum capacitance}

Transport in TI's have been studied extensively in comparison to its electrostatic properties, but the latter form an integral part of a comprehensive understanding of their electronic properties. Quantum capacitance measurements, especially in doped TI's~\cite{doped1,doped2}, can reveal the nature of the temperature dependent DOS and enhance our understanding of their applicability to spintronics~\cite{SpinA1}. Such measurements have been reported on carbon nanotubes and mono- and bilayer graphene systems~\cite{nano1,nano2} and hold sway on the future of device physics, and can lead to performance enhanced field effect transistors~\cite{FET}. This section is dedicated to the discussion of the quantum capacitance in an ultra-thin TI at zero temperature. While our formalism does not incorporate a temperature dependence, it is unique in that it accounts for the effects of hexagonal warping.
\\
\\
Quantum capacitance ($C_{Q}$) is readily defined as 
\beq C_{Q} = e\frac{\partial Q}{\partial E_{F}} = e^{2}D_{T},\eeq where $E_{F}$ the Fermi energy and $D_{T}$ is the temperature dependent DOS, and shown below is a standard form for it.
\beq D_{T} = \int_{0}^{\infty}dE\frac{\partial f(E-E_{F})}{\partial E_{F}},\eeq Here, $f(E)$ is the Fermi-Dirac distribution function, and in the limit of zero temperature, $D_{T=0} = D(E_{F})$, since the Fermi distribution approaches a Heaviside Step function. Refering to $E^2_{n s}$ (the index $\alpha$ is set to 1) as $\Delta E_n$ in order to avoid ambiguity, it is easy to write the expression of $D(E_{F})$ from equation (\ref{22}) as 
\beq D(E_{F}) = \sum_{s= \pm 1}  \frac{{D_0}(E)}{2} \left[ \frac{1 - exp\{-(\frac{2\pi \Gamma_0 E}{\hbar^2  w_c^2})^2\}Cos(\pi \frac{[{E}^2 - (\Delta_z + s \Delta_t)^2]}{\hbar^2 w_c^2})}{1 - 2exp\{-(\frac{2\pi \Gamma_0 E}{\hbar^2  w_c^2})^2\}Cos(\pi \frac{[{E}^2 - (\Delta_z + s \Delta_t)^2]}{\hbar^2 w_c^2}) + exp\{-2(\frac{2\pi \Gamma_0 E}{\hbar^2  w_c^2})^2\}}\right] \left(1-\frac{ E^2_{n_{0}}}{(\hbar w_{c})^{2}}\right) = \frac{C_Q}{e^2}\eeq
By appropriate substitution for $E_{F~n_0}$, a complete expression for zero temperature quantum capacitance can be obtained.
\\
\begin{figure}[ht]
\centering
\begin{minipage}[b]{0.45\linewidth}
\fbox{\includegraphics[scale=0.64]{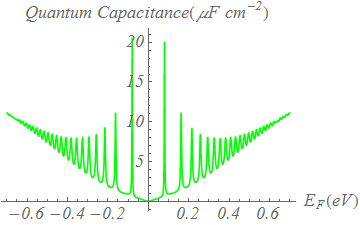}}
\caption{A plot of quantum capacitance with the Fermi energy depicting the oscillations around the point of charge neutrality, which decrease in amplitude far away from $E_F = 0$.}
\label{fig:minipage1}
\end{minipage}
\quad
\begin{minipage}[b]{0.45\linewidth}
\fbox{\includegraphics[scale=0.62]{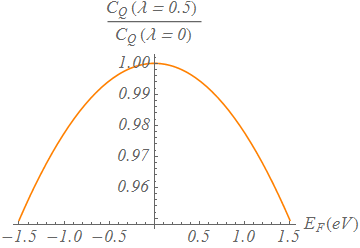}}
\caption{This shows the ratio of quantum capacitances at warping parameter values $\lambda = 0.5$ and $\lambda =0$ versus Fermi Energy}
\label{fig:minipage1}
\end{minipage}
\end{figure}

\begin{figure}[ht]
\centering
\begin{minipage}[b]{0.45\linewidth}
\fbox{\includegraphics[scale=0.63]{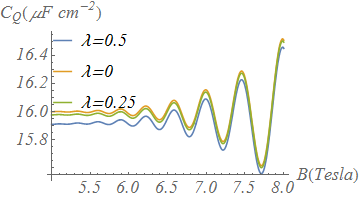}}
\caption{A plot of quantum capacitance with the magnetic field for different values of the warping parameter.}
\label{fig:minipage2}
\end{minipage}
\quad
\begin{minipage}[b]{0.45\linewidth}
\fbox{\includegraphics[scale=0.38]{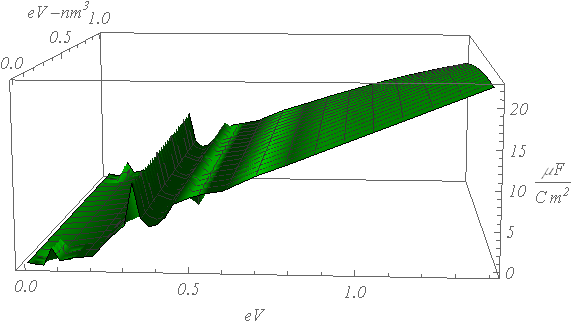}}
\caption{This shows the variation of quantum capacitance with warping parameter and Fermi energy.}
\label{fig:minipage2}
\end{minipage}
\end{figure}

In Fig.4 we can see oscillations in quantum capacitance with consistent~\cite{Firoz1, Tahir} but additional amplitude damping as a contribution from the hexagonal warping term, which is small due to the perturbative nature of $\lambda$. This is demonstrated in Fig.5 where the amplitude damping at $\lambda = 0.5$ eV-nm$^3$  is compared to the case when warping is absent by plotting the ratio of quantum capacitances at these values of $\lambda$. The form of the DOS and consequentially that of quantum capacitance is asymptotic, and it fails to recover the electron-hole asymmetry at the point of charge-neutrality. However, it proves useful in predicting the behavior of the quantum capacitance for larger values of Fermi energy and shows a correspondingly receding envelope of these oscillations. 

SdH oscillations are observed in Fig.6 with known  amplitude variation~\cite{Tahir} for different values of the warping parameter, and with reduced average capacitance and frequency damping as the warping contribution rises, the latter of which may be due to the asymptotic form of the DOS. Fig.7 shows the variation of $C_Q$ with $E_F$ and $\lambda$, and we see the quantum capacitance increasing with Fermi Energy due to charge accumulation. For a given $E_F$, we notice that the warping term increments lead to a dip in quantum capacitance which is consistent with our earlier results.

The ratio of values of quantum capacitance (Fig.5) are of the order of $10^{-2}$ at magnetic field of 5T, and there is little change to this ratio on variation of the magnetic field in the range 1T and 10T. A measurement of this ratio upto the stated order, along with observations of reduced frequency SdH oscillations, and receding oscillations in quantum capacitance as a function of Fermi energy in different materials such as $Bi_{2}Te_{3}$ and $Sb_{2}Te_{3}$ with experimentally established warping parameters ~\cite{Fu} in field strength ranges of 2T to 8T  might provide experimental proof of the results stated here.

\section{conclusion}
In this paper, we have investigated the properties of a warped ultra-thin TI film in the presence of a perpendicular magnetic field. In the ultra-thin limit, hybridization between the top and bottom surfaces of the TI is considered, along with the usual relativistic Dirac piece and the term due to Zeeman splitting. In addition and significantly, we've managed to incorporate the effects of the newly proposed hexagonal warping component by treating it perturbatively, and have obtained the correction to the LL's, demonstrating that the first order energy correction vanishes. The second order correction is readily obtained and is shown to be miniscale in comparison to the unperturbed LL's, thereby justifying our methods.

Calculations for the DOS is done analytically and leads to the recovery of an expression found in a recent publication~\cite{Firoz1}. This form is coupled with a factor accounting for the warping contribution and provides a complete analytical expressions for the DOS for an ultra-tin TI film. 

Furthermore, the Hall conductivity of the warped TI film is derived by determining the first order state corrections and then employing the Kubo formula. Once again, we recover established results~\cite{Burk}, including but not limited to a quantum phase transition, and go on to show that the warping term has little effect on it, making this a useful observation for 2D TI transport. 

Lastly, we've turned to the calculation of zero-temperature quantum capacitance, using it's very direct connection to the Fermi surface DOS. In the process of our analysis, we found that quantum capacitance persists with it's show of oscillations as a function of Fermi energy but they are further damped  due to the presence of the warping term and we predict that these oscillations tend to die out for large values of $E_F$. SdH oscillations are also observed with remarkable similarity to established results~\cite{Tahir} but with a reduced average capacitance and damped frequency. 

While the effect of quantum capacitance can be ignored in most conventional devices, it is crucial to the development of nano scale technology which employs ultrathin topological insulators. To this end, it is important to note that while experiments have found hexagonal warping, there have been no experiments which have measured the effects of hexagonal warping on quantum capacitance in ultrathin TI's. Experiments have measured Hall conductivity and quantum capacitance~\cite{Stan,Xiu}, wherein the sample was thick enough to minimize the effects of hybridization of the top and bottom surfaces, and without considering the effects of hexagonal warping. We've shown that our results are consistent with the findings of such experiments and the corresponding theoretical analyses.  As such, one can measure the quantum capacitance by varying the channel voltage between a gate and an ultrathin TI, and noting the corresponding charge response. We propose that future experiments be conducted on gated devices in conjunction with ultrathin TI's, with parametric ranges consistent with those used in this paper, the results of which may be used to validate our claims.

During this discourse, we have  stressed on the derivations of quantities fundamental to almost any condensed matter system, both in transport and in electrostatics, and stated possible ranges of parameters to facilitate experimental validation. It is felt that significant contributions in the understanding of thin-film TI's lie in the understanding of the hexagonal warping and the results presented here may be adopted to analyse other interesting properties of ultrathin topological insulators. \\

{\bf Acknowledgement}:
Debashree Chowdhury would like to acknowledge DST (DST/PHY/20120242) for financial support.

\end{document}